\documentclass[preprint]{aastex}

\begin{document}

\title{Superluminous X-rays from a Superluminous Supernova}

\shorttitle{X-rays from a Superluminous SN}

\author{A.~J. Levan\altaffilmark{1}, A.M. Read$^{2}$, B.D. Metzger$^{3}$, P.J. Wheatley$^{1}$, N.R. Tanvir$^{2}$ }

\email{a.j.levan@warwick.ac.uk}

\altaffiltext{1}{Department of Physics, University of Warwick,
  Coventry, CV4 7AL, UK } 

\altaffiltext{2}{Department of Physics and Astronomy, University of
  Leicester, University Road, Leicester, LE1 7RH, UK }

\altaffiltext{3}{Department of Physics, Columbia University, 538 West 120th Street, 704 Pupin Hall,
MC 5255, New York, NY 10027, USA. }


\begin{abstract}
The discovery of a population of superluminous supernovae (SLSNe), with peak luminosities 
a factor of $\sim 100$ brighter than normal SNe (typically SLSNe have
$M_V <-21$), has shown an unexpected diversity in core-collapse supernova properties. 
Numerous models have been postulated for the nature of these events, including 
a strong interaction of the shockwave with a dense circumstellar environment, a
re-energizing of the outflow via a central engine, or an origin in the catastrophic 
destruction of the star following a loss of pressure due to pair production in an
extremely massive stellar core (so-called pair instability supernovae). 
Here we consider constraints
that can be placed on the explosion mechanism of
Hydrogen-poor SLSNe (SLSNe-I) via X-ray observations, with {\em XMM-Newton, Chandra} and
{\em Swift},  and show that
at least one SLSNe-I is likely the brightest X-ray supernovae ever observed, with
$L_X \sim 10^{45}$ ergs s$^{-1}$, $\sim$150 days after its initial discovery. This is a
luminosity 3 orders of magnitude higher than seen in other X-ray supernovae powered
via circumstellar interactions. 
Such high X-ray luminosities are sufficient to ionize the ejecta and markedly
reduce the optical depth, making it possible to see deep into the ejecta
and any source of emission that resides there. Alternatively,
an engine could have powered a moderately relativistic jet external to
the ejecta, similar to those seen in gamma-ray
bursts. If the detection of X-rays does require an engine it implies that
these SNe do create compact objects, and that the stars are not completely destroyed in a
pair instability event. Future observations will determine which, if any, of
these mechanisms are at play in superluminous supernovae.

\end{abstract}



\section{Introduction}

The past decade has seen a major improvement in our ability to locate and track 
transient events, largely thanks to numerous time-resolved, wide and often deep sky surveys, combined with flexible and capable multiwavelength followup \citep[e.g.][]{rau09,drake12}. A striking success of these campaigns has been the revealing of a 
broad diversity of cosmic explosions from final moments in the lives of massive stars. At the bright end, long
duration gamma-ray bursts (GRBs) have been successfully tied to stripped core-collapse
type Ic supernovae \citep[e.g.][]{hjorth11}, and can reach peak absolute magnitudes in excess of $M_B < -38$ \citep{racusin08,bloom09}. At the
faint end, the discovery of transients with peak magnitudes of $M_V \sim -10$ may
be suggestive of  either electron capture SN \citep[e.g.][]{nomoto87}, or outbursts from stars in the final stages of
their lives \citep[e.g.,][]{maund}. The dynamic range in luminosity 
between these two extremes is in excess of $10^9$, and highlights
the varied signatures of stellar death that are now being uncovered. 

Perhaps one of the most remarkable developments of the past few years has been
the discovery and characterisation of a population of superluminous SNe (SLSNe), whose
peak luminosities are in excess of $M_V < -21$ \citep{smith07,gal-yam09,quimby11},
see \citet{galyam12} for a recent review. 
Taken at face value these
peak luminosities suggest the synthesis of several $M_{\odot}$ of $^{56}$Ni during the
explosion, a mass which is implausible for anything but the most massive stars. This led
to the suggestion that these SNe may represent the long sought after pair instability 
SNe (PISNe) \citep[e.g.][and references therein]{quimby07,woosley07,galyam12}, with
SN~2007bi representing a particularly promising example \citep{gal-yam09}.
In these supernovae, pair production in the core of the massive star drastically
reduces the central pressure, creating a run-away collapse followed by detonation \citep[e.g.][]{rakavy67}.
These SNe are of particular importance, since the build up of a sufficiently massive stellar core
is likely only to be possible in particularly massive and low metallicity stars \citep[e.g.,][]{heger2003}. 
Depending on the details of mass loss and rotational mixing,
this may make pair instability a dominant channel for the collapse of first generation, population III stars, making their
identification and study of significant import for our understanding of physical 
processes in the early Universe, including the initial enrichment of the IGM, the
contribution of stars to reionization, and the setting of the mass function
of future generations of stars.  

However, the interpretation of SLSNe as pair instability events is far from unambiguous based on the available data, 
and two widely discussed alternative models have been suggested. In
the first the shock
wave from the SN is re-energized by an engine inside the supernova, commonly
considered to be either a magnetar tapping the rotational energy of
the neutron star \citep{kasen10}, or an accreting black hole \citep{quataert12,Dexter&Kasen12}.
Alternatively, in the second model the luminosity is created via an unusually strong circumstellar
interaction, perhaps due to large scale mass loss from the progenitor in the decades prior 
to its collapse \citep{chevalier11}.

In practice, the properties of SLSNe are also varied, in particular they appear to divide into 
broad classes \citep{galyam12}. These split between SLSN II, hydrogen rich explosions, often with strong narrow
H-lines \citep[e.g.][]{smith07}, and extreme SN Ic -- hydrogen poor explosions, with very atypical spectral 
features \citep{barbary09,soker10,quimby11}. In addition, a population of SNe showing clear exponential tails due to Nickel 
production (SLSN-R) may represent the best candidates to be PISN-like
events \citep[e.g.][]{gal-yam09,galyam12}. 

To date, the majority of diagnostics of the nature of these exceptionally luminous
SNe have focussed on longer wavelengths regimes, unsurprisingly since this is where
their luminosities seem to peak. Here we consider an alternative approach of utilizing 
the insights that can be gained from
sensitive X-ray observations to distinguish between plausible models. We present
a thorough re-analysis of X-ray observations of the first identified SLSN-I, SCP 06F6, 
confirming it to be a luminous X-ray emitter at late times after the initial identification of
the outburst, if this X-ray emitting phase was at all extended then the 
X-ray energy output could rival, or exceed that seen in the optical. 
We additionally show that limits for other SLSN-I observed in the X-ray band, 
imply that such X-ray emission is rare ($<10$\% of SLSN-I), and consider
models that can account for such rare, but luminous X-ray emission.

\section{X-rays from SCP 06F6}
\label{scp06f6}
SCP 06F6 marks a proto-type of the SLSN-I events, although the
properties of this event were sufficiently unusual that
it initially defied classification \citep{barbary09,gaensicke09,soker10}. Eventually, 
through the discovery of other similar events, it was recognized as a highly luminous supernova at $z=1.189$, 
with a peak magnitude brighter than M$_V$ = -22 \citep{quimby11}. Perhaps even more remarkable was the detection of 
SCP 06F6 in X-rays with {\em XMM-Newton} approximately 150 days after its first discovery
\citep{gaensicke09}. 
While this work utilized an incorrect redshift estimate for SCP 06F6, before its more accurate
identification, it already implied a high luminosity event. Corrected for the now known redshift
the X-ray luminosity becomes even more extreme, exceeding $L_X \sim 10^{45}$ ergs s$^{-1}$.

The {\em XMM-Newton} observations were obtained on 2 August 2006, 162 days after the
initial detection of the SCP 06F6 outburst, and significantly after the optical peak. A total
of 9916 and 8220 seconds of good exposure were obtained with MOS1 and MOS2 respectively and
2622 ks with the EPIC pn (a smaller total integration due to the longer
setup time, and greater sensitivity to background flaring).  The MOS1 observations were
obtained with the medium filter in place as a safeguard against any optical loading
of CCD by stars in the field (although the source is well out of the Galactic plane, and has
only one star brighter than R$\sim 12$ in the EPIC field of view, offset
several arcminutes from the position of SCP 06F6). The MOS2 and pn
observations were obtained with the thin filter. 

\begin{figure}
\includegraphics[width=\columnwidth,angle=90,scale=0.33]{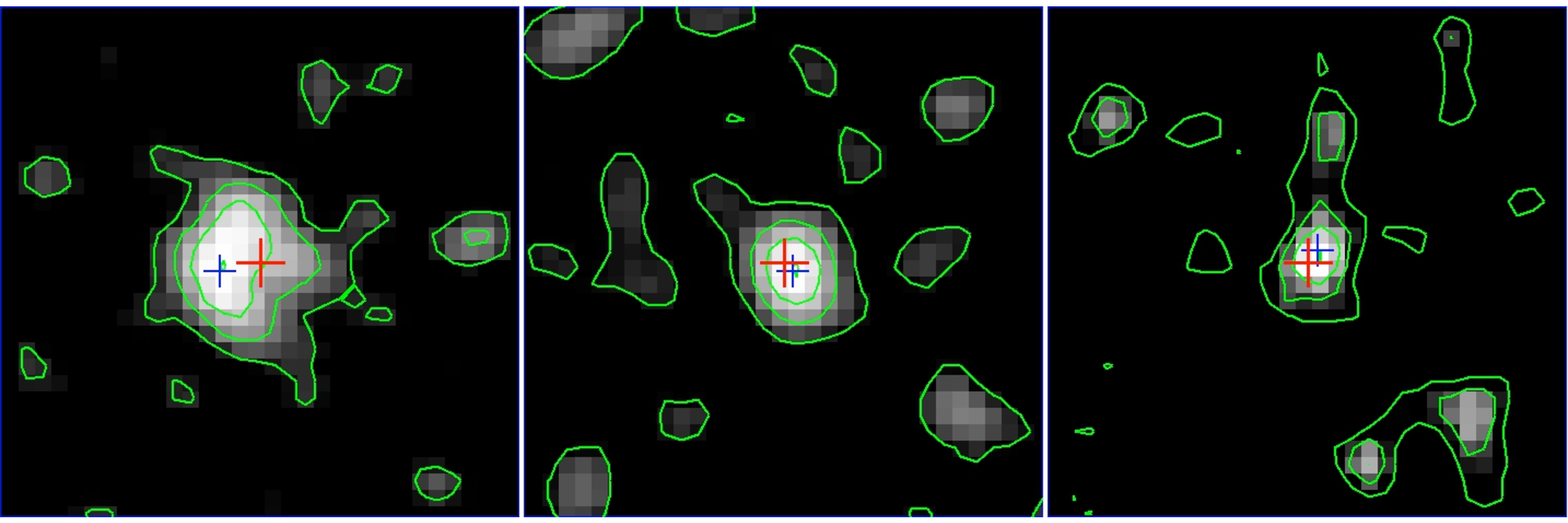}
\caption{\label{mos_im} Soft band images of SCP 06F6, obtained from XMM-Newton observations in 
the MOS1 (left), MOS2 (middle) and pn (right) detectors (MOS1 and MOS in 0.2-2 keV, and pn in 0.3-1 keV). 
Each image is approximately 2 arcminutes across. 
The source is clearly visible 
in MOS1 and MOS2, and detected at $\sim 3.0\sigma$ confidence in the pn. The red crosses
show the optical position of SCP 06F6 from \citep{barbary09}, while the blue crosses show
the positions derived for the X-ray source via the standard detection algorithms. }
\end{figure}

The {\em XMM-Newton} observations were significantly impacted by 
background flaring. Because of this we chose to restrict the energy band considered to
the 0.2-2 keV range and the extracted MOS images are shown in Figure~\ref{mos_im}. 
Within these images we performed source detection utilizing the
standard {\em XMM-Newton} SAS tools (version 11.0.0). In both the MOS1 and 
MOS2 images this returns a clear source, with count rates of $(6.5 \pm 1.4) \times 10^{-3}$ and
$(9.3 \pm 1.7) \times 10^{-3}$ cps respectively. The signal to noise ratio of these detections
is 4.7 and 5.3, while the reported detection likelihoods for
each source ({\tt DET\_ML}) are 18.7 and 24.3, corresponding to false positive 
probabilities of $8 \times 10^{-9}$ and $3 \times 10^{-11}$. In other
words, these detections are highly significant in each detector, especially
given the narrow spatial window searched (i.e. the false positive probabilities
are close to those above, since we have a single trial). 

The pn detectors are significantly more noisy, due to the shorter 
total exposure, and greater sensitivity to background flaring. Nonetheless there is a significant excess flux at the position of
SCP 06F6, with a measured rate of $0.0111 \pm 0.0037$ cps (0.3-1 keV, see Figure~1). For the
source in the pn {\tt DET\_ML} = 5.96, corresponding
to a false positive probability of 0.0025 (i.e. the source is detected
at $>3 \sigma$). Since the
optical paths for each of the detectors are different this suggests that we have three
independent detections of X-ray emission from SCP 06F6. 

The positions of the sources in the MOS1 and MOS2 images are 
RA= 14$^h$32$^m$28.18$^s$, DEC= 33$^d$32\arcmin 22.8\arcsec\, and RA= 14$^h$32$^m$27.22$^s$,  DEC= 33$^d$32\arcmin 22.9\arcsec, with reported centroid
errors of $\sim 3$ arcseconds on each position (with an additional systematic of 1-2\arcsec, each at the 1-$\sigma$ level). 
These positions are offset $\sim$10\arcsec~ and $\sim$2\arcsec~ respectively from
the position of the optical source reported by \cite{barbary09}. Although these errors are
relatively large the faintness of the source, and extended structure in the high background make
precise centroiding difficult. We note that a second X-ray source at 
RA= 14$^h$32$^m$46.20$^s$ DEC= 33$^d$38\arcmin 00.2\arcsec~ 
shows a similar offset
between MOS1 \& MOS2. SCP 06F6 is point-like in each exposure, and not consistent with
any known features in the {\em XMM-Newton} background during flaring intervals. 

\begin{figure}
\includegraphics[width=\columnwidth]{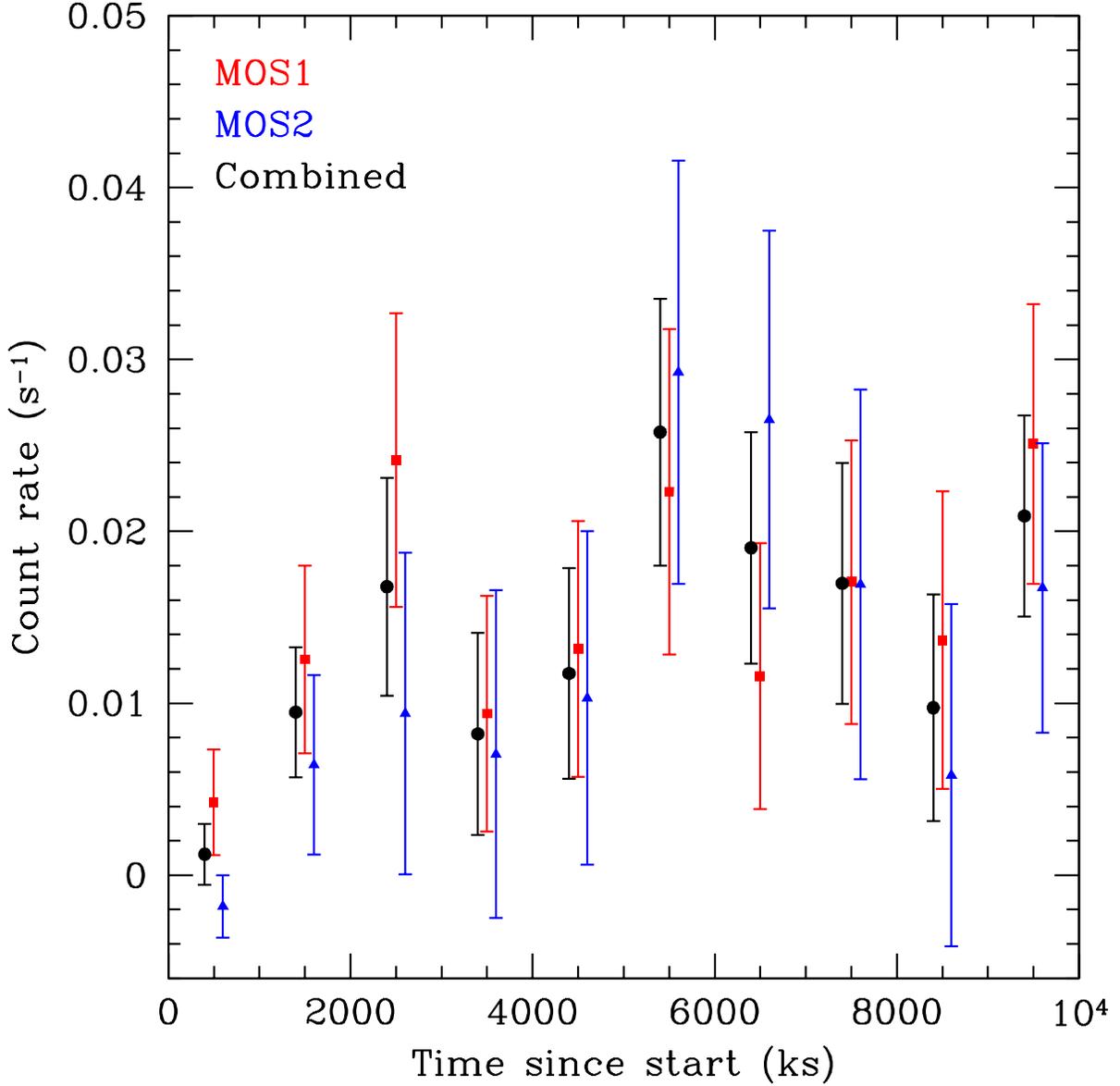}
\caption{\label{mos_lc} The MOS-1 (blue) and MOS-2 (red) lightcurves
of SCP 06F6. While the overall count rate is low, these lightcurves are suggestive
of variability in the source over the duration of the observation. The
$\chi^2/dof$ for a constant source is 1.56 and 2.70 for MOS1 and MOS2 respectively, and
is 3.86 for the combined lightcurve. This implies some
degree of variability which is driven by the first bin in each lightcurve. Note the times for each detector are identical, but have
been slightly offset for clarity.  }
\end{figure}

EPIC MOS (0.2-2 keV; PATTERN$<=$12) X-ray lightcurves in 1 ks bins were
extracted from 28 arcsec circles centred on the best optical position.
Equivalent background lightcurves were extracted from a 60-180 arcsec
annulus, centred on the same position.
The  X-ray lightcurves in the MOS1 and MOS2 detectors are shown in Figure~\ref{mos_lc}. 
While the total signal is weak, it is possible to split the 10 ks observations into 1 ks bins. 
 We fit the resulting light curves with a constant source, allowing only the normalization to vary,
the returned $\chi^2/dof$ for in these two observations are 1.56 (12.50/8) and 2.70 (21.58/8). This is suggestive of a degree of 
variability over the timescale of the observations.  Further, we also average the two lightcurves together, with 
their errors added in quadrature, for the resulting 
MOS1+MOS2 lightcurve, a constant source yields $\chi^2/$dof= 3.86 (30.90/8). These $\chi^2/$dof values indicate
probabilities for the constant model of 0.13, 0.006 and $10^{-4}$ for MOS1, MOS2 and the combined lightcurve
respectively. However, we also note that this poor fit is driven predominantly by the first point within the 
lightcurve, which is low in both MOS1 and MOS2. Ignoring this point would yield $\chi^2/$dof of 4.45/7 =0.64, 6.10/7 = 0.87 and
7.58/7 = 1.08 for each of MOS1, MOS2 and combined, these have probabilities of 0.73, 0.53 and 0.37 respectively, and hence
after the first $\sim$ 1000 s we would not consider the source to be variable. Similar changes in $\chi^2$ are not observed
for the removal of any other (randomly chosen) time bin. We therefore conclude that the variability
is driven by an apparent relatively rapid rise in flux within the first 1000 s of the observation, which is seen in both the
MOS1 and MOS2 detectors. Given the similarity of the lightcurves we do not believe this to be spurious, but since it is driven 
predominantly by these two points it should be treated with caution.

We extracted spectra of the source in the same apertures described above, utilizing 
ancillary response files (ARFs) and response matrix files (RMFs) created using the
point spread function model {\tt ELLBETA}, and correctly accounting for bad pixels. 
The resulting spectra are inevitably of low signal to noise, but do allow some
constraints on the spectral properties to be derived. Given the weakness of the source
we fix  the H-column density at the Galactic value ($N_{H({\rm Gal)}} = 8.85 \times 10^{19}$ cm$^{-2}$, \citet{dickey}). 
The resulting spectra are adequately fit using Cash-statistics giving either
$\Gamma=2.64^{+0.49}_{-0.36}$ for a power-law, or 
$kT = 1.55^{+0.69}_{-0.59} $, $Z=0.31^{+1.44}_{-0.31}$, for a {\tt mekal} thermal model (errors at 90\% confidence).
The spectral fits are of comparable quality for each model, with the C-stat and degrees of freedom being
913.91/768 and 915.08/767 for power-law and {\tt mekal} models respectively. 
Both the power-law slope and inferred temperature are typical of many astrophysical 
sources that may be of interest to the origin of SCP 06F6
such as SNe shock interactions \citep[e.g.][]{immler07} 
and GRB X-ray afterglows \citep{evans10,margutti13}. Hence, the spectra, while indicative of a relatively soft source, cannot 
provide a clear indication of the nature of the emission. 

Utilizing the above spectral models provides a 0.2-10 keV flux (corrected for 
foreground absorption) in the power-law or thermal
cases of  $F_{\Gamma} \approx 1.3\times 10^{-13} $ ergs s$^{-1}$ cm$^{-2}$ and $F_{kT} \approx 9 \times 10^{-14}$
ergs s$^{-1}$ cm$^{-2}$ respectively. The resulting k-corrected luminosities at $z=1.189$ are
$L_{X,\Gamma} = 1.8 \times 10^{45}$ ergs s$^{-1}$ and $L_{X,kT} = 6.5 \times 10^{44}$ ergs s$^{-1}$,
an extremely high luminosity in either case.  

SCP 06F6 was also observed by the {\em Chandra} X-ray observatory on 2006-11-04, with
the source placed on the S3 chip, and with an exposure time of 5ks. In a 5 arcsecond
aperture centred on the optical position of SCP 06F6 we find no counts in a 0.2-10 keV image, 
corresponding to a 3 $\sigma$ limiting count rate of $9 \times 10^{-4}$ cps, utilizing the method
of \cite{kraft}. There are also no sources detected within the 28\arcsec\, radius aperture utilized for
the {\em XMM-Newton} observations, suggesting there is not a contaminating unresolved source
observed by {\em XMM-Newton}.
For a simple $\Gamma=2.6$ model (as inferred from the {\em XMM-Newton} observations), 
$N_{H({\rm Gal})} =8.85 \times 10^{19}$ cm$^{-2}$, 
this implies a flux limit of $<1.4 \times 10^{-14}$ ergs s$^{-1}$ cm$^{-2}$, corresponding to a luminosity
of $< 2.5 \times 10^{44}$ ergs s$^{-1}$, a factor of five fainter than the {\em XMM-Newton} detections. 
This demonstrates that the source detected by {\em XMM-Newton} is transient. 
These observations and limits are compared with other transient X-ray
sources in Figure~\ref{lccomp}.

Sources with flux $\sim 10^{-13}$ ergs s$^{-1}$ cm$^{-2}$ are relatively rare on the sky, and a
typical log(N) - log(S) relation would predict only $\sim 100$ such sources per square degree
\citep{manners03,mateos08} in a similar (but not identical) 0.5 - 8 keV energy range. 
Hence the expected number within 10\arcsec~of a position
on the sky is $\sim 10^{-3}$ implying a low probability that the source is unrelated to 
SCP 06F6, especially when one also considers the temporal variations of a factor 
five in flux over $\sim 100$ days. We also
note that deep optical observations of the source by \cite{barbary09} show that any
host galaxy must have $z > 26.1(AB)$, corresponding to $M_B > -18$ at $z=1.189$, and
ruling out any bright AGN at the location of SCP 06F6. 
Therefore we conclude both that the {\em XMM-Newton} observations
detected a source at high confidence, and that this detection was associated with SCP 06F6. 

The inferred spectral energy distribution (SED) of the source at late times is shown in Figure~\ref{sed}.
The lack of contemporaneous X-ray 
and optical data makes drawing strong conclusions difficult. However,  we note that in the absence
of significant optical brightening over the $\sim 50$ days between the final epoch of 
optical observations and those with {\em XMM-Newton} we can conclude that i) the X-ray
luminosity is significantly in excess of the optical luminosity and ii) the extrapolation of the
X-ray spectral slope
(if interpreted as a power-law) lies above the likely optical luminosity. 
We note that
the source is not detected in the {\em XMM-Newton} optical monitor observations. The
depth of these observations rules out the brightest possible extrapolation of the
X-ray flux, but in general is not highly constraining. 

\section{X-ray observations of other SLSNe}
Motivated by the detection of SCP 06F6 in X-ray's we have searched for other 
SLSNe with X-ray observations. Numerous SLSNe have been observed by 
{\em Swift} to obtain both UV and X-ray observations, and here we consider constraints
that can be placed on the hydrogen-poor SLSN-I, with our sample shown in Table 1. 

For each of these SNe we extracted XRT images over the 0.3-10 keV band for each observation, 
as well as combining multiple observations where possible to obtain deeper constraints. 
We measured the observed XRT counts in each image in apertures of 5 pixel radius (11.8 arcseconds), and
measured the background in a large (100 pixel radius) background region,  at a location close to the centre of the chip, but free of obvious
bright X-ray sources. We corrected the measured count rates for the impact of bad pixels and masked out columns within the XRT using the exposure maps for each snapshot observation \citep[e.g.][]{evans09}, which reduce the effective on-source exposure time in some snapshots. We assessed the significance of the resulting number counts with reference to 
the Bayesian method of \citet{kraft}, providing upper limits at the 99\% confidence level. We then corrected this flux (or limiting flux) for
the limited PSF contained within the aperture radius, using the published enclosed energy curves. From this we 
converted the count rate
to physical fluxes utilizing a generic spectral model (Photon index $\Gamma=2$,
$N_H$ (Galactic) was set according to each SNe from the Galactic models of \cite{dickey}). We note
that several of the SNe in our sample have also been presented in \cite{ofek12}. These authors
find limiting luminosities which are typically a factor of $\sim 3$ brighter than those
presented here, although this difference is largely explained by the differing choice
of spectral index ($\Gamma = 0.2$ in \citet{ofek12} and $\Gamma=2$ in this work), 
which results in a difference in flux of a factor of $\sim 3$ at $z=0$

The supernovae for which such observations are available, along with the inferred fluxes
and luminosity limits are shown in Table~\ref{x-ray}. In our sample, besides our
initial observations of SCP 06F6, we only find a detection at the 99\% level for
a stacked image of CSS121015, at a luminosity of $2.9^{+2.32}_{-2.23} \times 10^{42}$ ergs s$^{-1}$. However, given the number of measurements made (72 individual snapshots, and 12
stacked observations) this is consistent with the expectations of random noise, and therefore
we have only one statistically significant detection of X-rays, from SCP 06F6\footnote{We note that
the detection of SCP 06F6 is at higher confidence level, suggesting that we wouldn't expect any 
such significant signals within our sample. Further, since these measurements represented the
initial detections from a single trial (i.e. our subsequent search was
motivated by their discovery), they are not subject to the same sample size concerns. }.
We note that there is a possible detection of the SLSN-II 2006gy
with {\em Chandra} \citep{smith07}, at a high, but not exceptional luminosity of $2 \times 10^{39}$ ergs
$s^{-1}$. 
The implied limiting luminosities of these SN are shown graphically in Figure~\ref{lccomp}. In many
cases they clearly lie well below the luminosity of SCP 06F6. 
This directly implies
that any luminous X-ray emission in SLSNe must either be transient on time scales much
shorter than the timescale of the optical outburst, such that the {\em Swift} observations
have failed to detect it, while the {\em XMM-Newton} observations were favourably timed; 
or, perhaps more likely, that X-ray
emission in SLSNe is rare.

\subsection{Stacked limits from the {\em Swift} XRT sample}
Most of the SLSNe considered here we observed with {\em Swift} over multiple epochs, it is therefore possible
to stack the individual observations of each object in order to obtain a deeper limit for each SNe, at the cost
of losing temporal information. These individual limits are also shown in Figure 3.
Furthermore, given the large number of observations (a total of 72 individual observations are reported in Table 1) it is also
possible to stack the resulting images to obtain a super-stack, providing a total of 168770s of exposure time. 
This image is shown in Figure~\ref{stack}, along with the expected position of any SN signal. In this image we 
observe a total of 15 counts in a 5 pixel aperture, with an expected background of 11. Again using the
method of \cite{kraft}, this suggests a total count rate $<1.4 \times 10^{-4}$ s$^{-1}$ (corrected for encircled energy), 
a mean X-ray flux of $F_X < 5\times 10^{-15}$ ergs s$^{-1}$ cm$^{-2}$. For a mean luminosity distance
of $d_L =1204$ Mpc ($\bar{z}=0.24$), this corresponds to an X-ray luminosity
of $L_X < 9 \times 10^{41}$ ergs s$^{-1}$. Since this luminosity represents a stacked limit for
12 SLSNe, the average X-ray luminosity for each event is $L_X < 7 \times 10^{40}$ ergs s$^{-1}$, although as
shown in Table 1, the results for each individual SNe vary widely due to the differences in exposure time and redshift. 
In other words, the average X-ray luminosity of an non X-ray detected SLSN-I is a factor of $<10^4$ fainter
than the inferred for SCP 06F6,  this suggests that SCP 06F6 is not simply towards the bright end of an 
essentially continuous luminosity function, and that is was much more luminous that most SLSNe. Indeed, 
these limits are approaching the brightness of the brightest
``normal" X-ray SN seen to date \citep[e.g.][]{dwarkadas} (see below).

\subsection{Comparison with other objects} 
Figure~\ref{x-ray} shows the X-ray luminosity of SCP 06F6 in context with other transient X-ray emitting sources, in particular X-ray supernovae,
gamma-ray bursts and candidate tidal disruption flares. The lightcurves for the latter sources are taken from
the {\em Swift} repository \citep{evans07,evans09}, with the bright GRBs 060729 \citep{grupe07,grupe10} and 080319B \citep{racusin08,tanvir10} are used to
show the extremely luminous side of the distribution and the unusual low luminosity GRBs 060218 \citep{campana06} and 100316D \citep{starling2011}
used to show the faint end. The candidate relativistic tidal flares are {\em Swift} J1644+57 \citep{levan11,bloom11,burrows11,zauderer12} and {\em Swift} J2058.4+0516
\citep{cenko12}. 
We also show numerous X-ray supernova detections of the past several years. These are
taken from \cite[][and references therein]{kouveliotou04,dwarkadas}, as well as the shock breakout
discovered for SN~2008D \citep{soderberg08}. 

SCP 06F6 lies towards the brighter end of this distribution. In particular, it is as bright as the
brightest GRB afterglows at the same point after core collapse, and more luminous by far
than any previously identified SNe. Interestingly, it is comparable in brightness to {\em Swift J1644+57} and {\em Swift} J2058+0516, which are generally thought to be related to tidal disruption events, 
but have also may be core collapse events \citep{quataert12,woosley12}. Below
we consider plausible physical mechanisms that could power such luminous X-ray emission. 

\section{Implications}

\subsection{The X-ray Emission Could Originate from Deep Within the Ejecta }
\label{sec:ionization}

We begin by describing how the observed X-ray emission from SCP 06F6 constrains the ionization state and opacity of the supernova ejecta and hence provides clues to its own origin.  

Assuming that the explosion responsible for SCP 06F6 expands homologously, the average density of the ejecta decreases with observer time $t$ as $\rho_{\rm ej} \simeq M_{\rm ej}/(4\pi/3 R_{\rm ej}^{3})$, where $M_{\rm ej}$, $R_{\rm ej} = v_{\rm ej}t/(1+z)$, and $v_{\rm ej}$ are the mass, radius, and velocity of the ejecta, respectively.  If the ejecta is nearly fully ionized (to be justified below), then the ionization state of a hydrogenic ion of charge $Z  = 26Z_{26}$ is determined by comparing the absorption rate of ionizing photons $\mathcal{R}_{\rm ion} = \mathcal{C}n_{\gamma, \nu > \nu_{1}}\sigma_{\nu_{1}}c$ (per ion) to the rate of recombination $\mathcal{R}_{\rm rec} =  n_{e}\alpha_{\rm rec}$, where $n_{\gamma, \nu > \nu_{1}} \approx L_{\nu_{1}}/4\pi h\nu_{1} R_{\rm ej}^{2}c$ is the number density of ionizing photons; $\mathcal{C}$ is a constant of order unity that depends on the spectrum of the ionizing radiation; $L_{\nu_{1}}$ is the specific X-ray luminosity near the ionization threshold energy $h \nu_{1} = {\rm Ryd}Z^{2} \simeq 10 Z_{26}^{2}$ keV; $\sigma_{\nu_{1}} \simeq 8\times 10^{-21} Z_{26}^{-2}$ cm$^{2}$ is the photoionization cross section at $\nu = \nu_{1}$; $n_{e} \simeq \rho_{\rm ej}/2m_{p}$ is the number density of electrons in the ejecta; $\alpha_{\rm rec} \approx 2.0\times 10^{-10}Z_{26}^{2}T_{4}^{-0.8}$ cm$^{3}$ s$^{-1}$ is the [type 2] recombination coefficient (e.g.~\citealt{Osterbrock&Ferland06}); and $T_{4} \sim 1$ is the temperature of the ejecta in units of 10$^{4}$ K.  Combining the above results, one finds that
\begin{eqnarray}
\frac{\mathcal{R}_{\rm ion}}{\mathcal{R}_{\rm rec}} \sim 5\left(\frac{L_{\nu_{1}}}{10^{44}{\rm erg\,s^{-1}}}\right)\left(\frac{M_{\rm ej}}{6 M_{\odot}}\right)\left(\frac{t}{150{\,\rm d}}\right)\left(\frac{v_{\rm ej}}{10^{4}{\,\rm km\,s^{-1}}}\right) Z_{26}^{-4}, 
\label{eq:ion}
\end{eqnarray}  
where we assume $\mathcal{C}, T_{4} \approx 1$ and have normalized $M_{\rm ej}$ and $v_{\rm ej}$ to values characteristic of the mass and velocity of the ejecta based on detailed models of the optical light curve of SCP 06F6 (\citealt{chatzopoulos09}).\footnote{Although the redshift of SCP 06F6 disagrees with those assumed by \citet{chatzopoulos09}, updated models accounting for the correct redshift give similar estimates (to within a factor $\lesssim$ 2) for the ejecta mass (M.~Chatzopoulos, private communication).}  We normalize $L_{\nu_{1}}$ to an estimate of the luminosity at the ionization frequency $\sim 10$ keV, if one were to extrapolate the observed X-ray emission to higher energy (assuming a power-law fit).   

Equation (\ref{eq:ion}) shows that on timescales $t \gtrsim 150$ days, the observed X-ray emission is sufficiently luminous to fully ionize even the K-shell electrons of Fe ($Z = 26$), i.e. $\mathcal{R}_{\rm ion}/\mathcal{R}_{\rm rec} \gtrsim 1$ for $Z \lesssim 26$.  This is even more true (i.e.~ $\mathcal{R}_{\rm ion}/\mathcal{R}_{\rm rec} \gg 1$) for lower-$Z$ elements in the frequency range $\sim 0.3-1$ keV of the observed X-rays.  Hence, by combining X-ray and optical observations, one can infer that the bulk of the supernova ejecta is most likely fully ionized during the observational window.

If the bulk of the ejecta is fully ionized, then its opacity is significantly lower than if the ejecta was partially neutral (as was likely the case earlier in its evolution) since the electron scattering opacity at soft X-ray frequencies is many orders of magnitude lower than the bound-free opacity.  The optical depth to Thomson scattering through the bulk of the ejecta is approximately given by
\begin{eqnarray}
\tau_{es} \approx n_{e}R_{\rm ej}\sigma_{T} \approx 10\left(\frac{M_{\rm ej}}{6M_{\odot}}\right)\left(\frac{t}{150{\,\rm d}}\right)^{-2}\left(\frac{v_{\rm ej}}{10^{4}{\,\rm km\,s^{-1}}}\right)^{-2}.
\label{eq:taues}
\end{eqnarray}
Equation (\ref{eq:taues}) (the use of which is justified by equation [\ref{eq:ion}]) shows that depending on the mass and velocity of the ejecta, the optical depth of the ejecta approaches $\sim$ few on timescales similar the observed X-ray emission.  This implies that the observed X-ray emission could in principle originate interior to the bulk of the ejecta.  Given these hints, we now discuss three possible origins for the X-ray emission from SCP 06F6.    

\subsection{Circumstellar Interaction}
The vast majority of X-ray luminous SNe result from the interaction of the outgoing shockwave with circumstellar material, emitted either continuously by the stellar wind of the progenitor, or in distinct mass loss episodes (e.g due to a Luminous Blue Variable phase).  The X-ray luminosity is a sensitive tracer of the mass loss rate, but even high mass loss rates of $\sim 10^{-4}$ M$_{\odot}$ yr$^{-1}$ only lead to an X-ray luminosity of $L_X \sim 10^{40}$ ergs s$^{-1}$. It has been suggested that both SLSN II and SLSN Ic may be due to strong CSM interaction, but in the case of SN~2006gy \citep[(IIn)][]{smith07}, the X-ray luminosity is orders of magnitude lower than necessary to power to optical display, leading \cite{chevalier12} to suggest that photoabsorption by the wind
itself effectively lowers the X-ray emission in these dense winds.  However, as discussed above in $\S$\ref{sec:ionization}, this is less of a concern in SCP 06F6 since the X-rays are so luminous that the bulk of the ejecta (and probably any surrounding CSM) is fully ionized.  Since photoabsorption is irrelevant in this case, even X-rays produced by the interaction of the bulk of the ejecta with CSM could in principle be directly observed.  Indeed, it is interesting to note that shocks created
in pulsational pair instability \citep[e.g.][]{woosley07} may at times create these conditions
allowing the X-rays created in shocks to escape if their luminosity is high enough to ionize the
medium. This may create a dichotomy in the X-ray luminosity seen from SNe, in which only
the highest X-ray luminosity can ionize the medium and hence become visible. 

Despite its promise, the CSM interaction model runs into several potential difficulties.  First, it appears difficult to explain the observed X-ray variability on the moderately short timescale of the observations ($t \lesssim 10^{4}$ s), since the light curve from CSM interaction would smeared out over the light crossing time of the bulk of the ejecta, $t_{\rm min} \sim R_{\rm ej}/c \sim 5(v_{\rm ej}/10^{4}{\rm km\,s^{-1}})(t/150{\rm\,days})$ days.  The model also appears to require some fine tuning, since in order to produce the observed luminosity a sufficiently large mass of CSM (at least several solar masses) must be concentrated in a radially thin shell on the radial scale $\sim 10^{16}$ cm, despite the lack of clear evidence for CSM interaction, e.g. emission lines, at earlier stages in the supernova.  Finally, all CSM interaction models are fundamentally limited by the kinetic energy of the explosion, while any X-ray emission persisting for long time scales at the observed luminosity $L_X > 10^{45}$ ergs s$^{-1}$ would radiate $10^{51-52}$ erg on a timescale of days to weeks. This is larger than the canonical kinetic energy released in most SNe, and would require an exceptionally powerful event, that was extremely efficient in converting kinetic energy to X-rays. Considerations of shock-created X-rays fail to achieve anything close to the luminosity of SCP 06F6 \citep[e.g.][]{ofek12},  although a recent paper has considered the case of SLSNe in more detail, and suggests that such high luminosities my be possible \citep{pan13}.
 
\subsection{Exposed Magnetized Nebula Powered by a Central Compact Object}

It has also been suggested that SLSNe could be powered by energy injection from a central compact object, such as a rapidly spinning, highly magnetic neutron star (a `millisecond magnetar') \citep{kasen10} or a newly-formed accreting black hole (e.g., \citealt{quataert12}, \citealt{Dexter&Kasen12}).  Although an extremely powerful jet from the compact object would puncture the stellar envelope (possibly producing a high energy transient such as a gamma-ray burst; see $\S\ref{sec:jet}$ below), a luminous SN could instead result if the jet power is lower, such that the energy in the relativistic outflow is trapped behind the supernova ejecta.  Particularly luminous optical emission is possible in this case if the compact object releases significant energy on a timescale $\sim$ weeks-months comparable to the photon diffusion time through the ejecta \citep{kasen10}.  

In such `engine powered SN' scenarios, the outflow from the compact object would produce a magnetized nebula (analagous to a scaled-up version of a pulsar wind nebula) inside the supernova ejecta (e.g.~\citealt{Bucciantini+07}).  If the ejecta mass is sufficiently low, and if energy input from the central compact object remains active until late times, then once the bulk of the ejecta becomes transparent to X-rays, radiation from the interior nebula could in principle escape to the observer.  As shown in equation (\ref{eq:taues}), this is plausible in the case for SCP 06F6 since the ejecta becomes transparent to X-rays ($\tau_{\rm es} \lesssim $ few) at $t \gtrsim 150$ days if the ejecta mass is reasonably low.  Interestingly, the X-ray luminosity of SCP 06F6 is of the same order of magnitude as (though somewhat higher than) its peak optical luminosity.  Comparable luminosities are in fact one prediction of such a model, if the power in the the outflow from the central compact object is approximately constant over the time interval from the optical peak emission to the X-ray observations ($t \sim 100-150$ days), a reasonable expectation.  

It is interesting to note that a generic feature of models containing a powerful central 
compact object is that they will act to ionize the ejecta, and so may make it optically thin 
much earlier than in cases where such an object is not present. This may provide a powerful
means of probing the nature of these SNe -- by directly searching for this emission -- assuming
it is active for time periods long enough to ionize the medium. i.e. that magnetar power,
or accretion in the case of a black hole, continues for several hundred days.  

Despite these merits, the `exposed nebula' model also runs into a few theoretical difficulties.  As in CSM interaction models, one would naively expect the X-ray variability to be limited to the light crossing time of the nebula, which is similar to that of the ejecta ($\sim$ several days on timescales or relevance).  The recent discovery of very rapid flaring from the Crab Nebula \citep{Abdo+11} has, however, illustrated that relativistic motion within the nebula may violate this constraint.  Another possible objection to such a scenario is the observed X-ray softness (low inferred $N_H$ local to the source), which appears inconsisent with a sightline that passes through relatively dense ejecta; note again, however, that photoabsorption is weak if the ejecta is indeed fully ionized.

\subsection{Jetted Emission}
\label{sec:jet}

\begin{figure}
\includegraphics[width=\columnwidth]{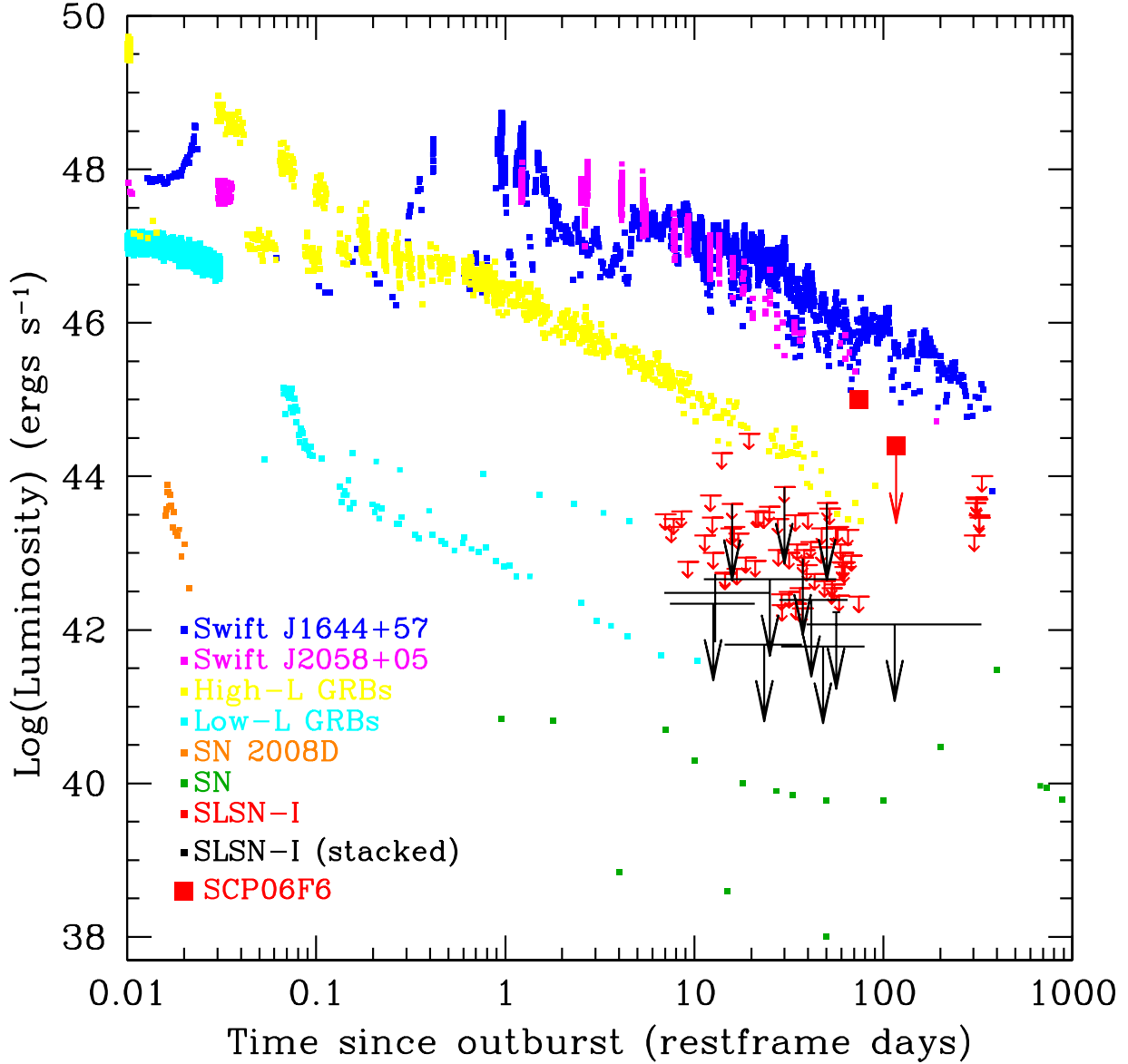}
\caption{\label{lccomp} The X-ray lightcurves (in luminosity space) of numerous extragalactic X-ray transient sources, including X-ray supernovae,
high- and low-luminosity GRBs, and candidate tidal disruption events ({\em Swift} J1644+57 and J2058+05). The 
X-ray luminosity of SCP 06F6, at late times after the supernova lies at the extreme end of 
the luminosity distribution, well beyond the most luminous X-ray supernova. The non-detection 
by {\em Chandra} approximately $\sim 100$ days later is suggestive of rapid variability, as
seen in some other sources \citep{levan11,saxton12}. In addition, the
plots shows limits inferred for other SLSNe from {\em Swift} XRT observations, as well as stacked limits to each
SNe (the extent of the horizontal line in each stacked limit indicates the time range of the observations). None 
of these show detections, implying either than X-ray emission such as that from SCP 06F6 is
rare, or that it occurs at much later times than the {\em Swift} observations are 
currently probing. Note: The times relative to outburst are relative to the trigger time
for GRB, and the earliest reported discovery date for the SNe, and therefore do not have
the same physical meaning, especially at early times (e.g. the SN are typically only
discovered a few days after core collapse). }
\end{figure}

\begin{figure}
\includegraphics[width=\columnwidth]{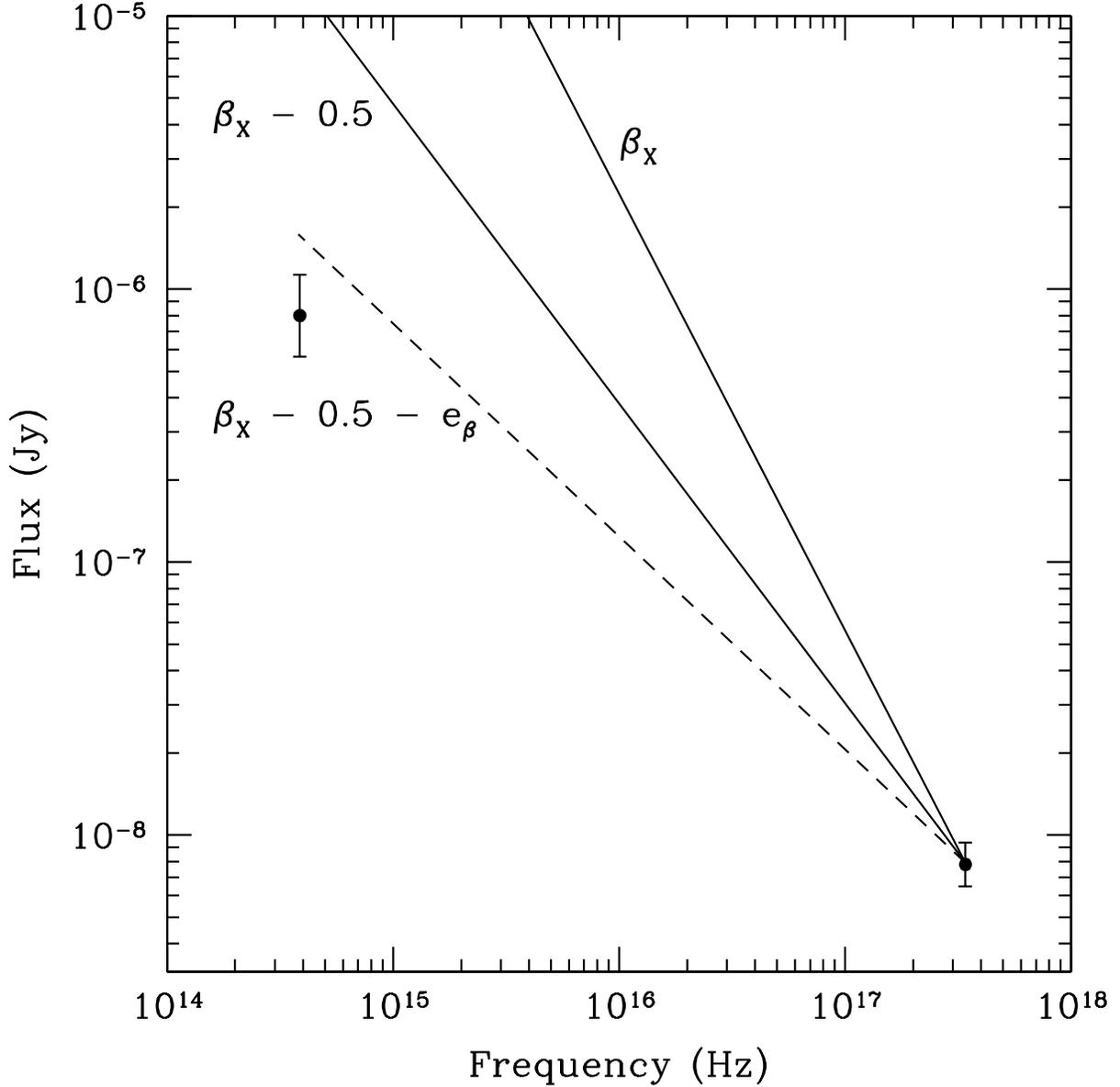}
\caption{\label{sed} The inferred late time spectral energy distribution of SCP 06F6 from
X-ray to optical wavelengths. The figure shows the X-ray flux (extrapolated from the X-ray spectrum), 160 days after 
the discovery of the outburst, 
compared with the optical flux at $\sim$100 days, and so the two are not simultaneous. However,
the extrapolation of the X-ray power-law (assuming the power-law model where $\beta_X = \Gamma-1$) is shown, and is 
substantially above the optical luminosity. The difference assuming the presence of a cooling
break between the two band is also shown ($\beta_X - 0.5$) as is an extreme example including
the 90\% error on the measured spectral slope, which, given the difference in times is
broadly marginally consistent with the optical flux. It should be noted that the X-ray luminosity
is significantly higher than the optical throughout, in contrast to other X-ray detected SNe \citep{ofek12}. }
\end{figure}

\begin{figure}
\begin{center}
\includegraphics[width=12cm,angle=270]{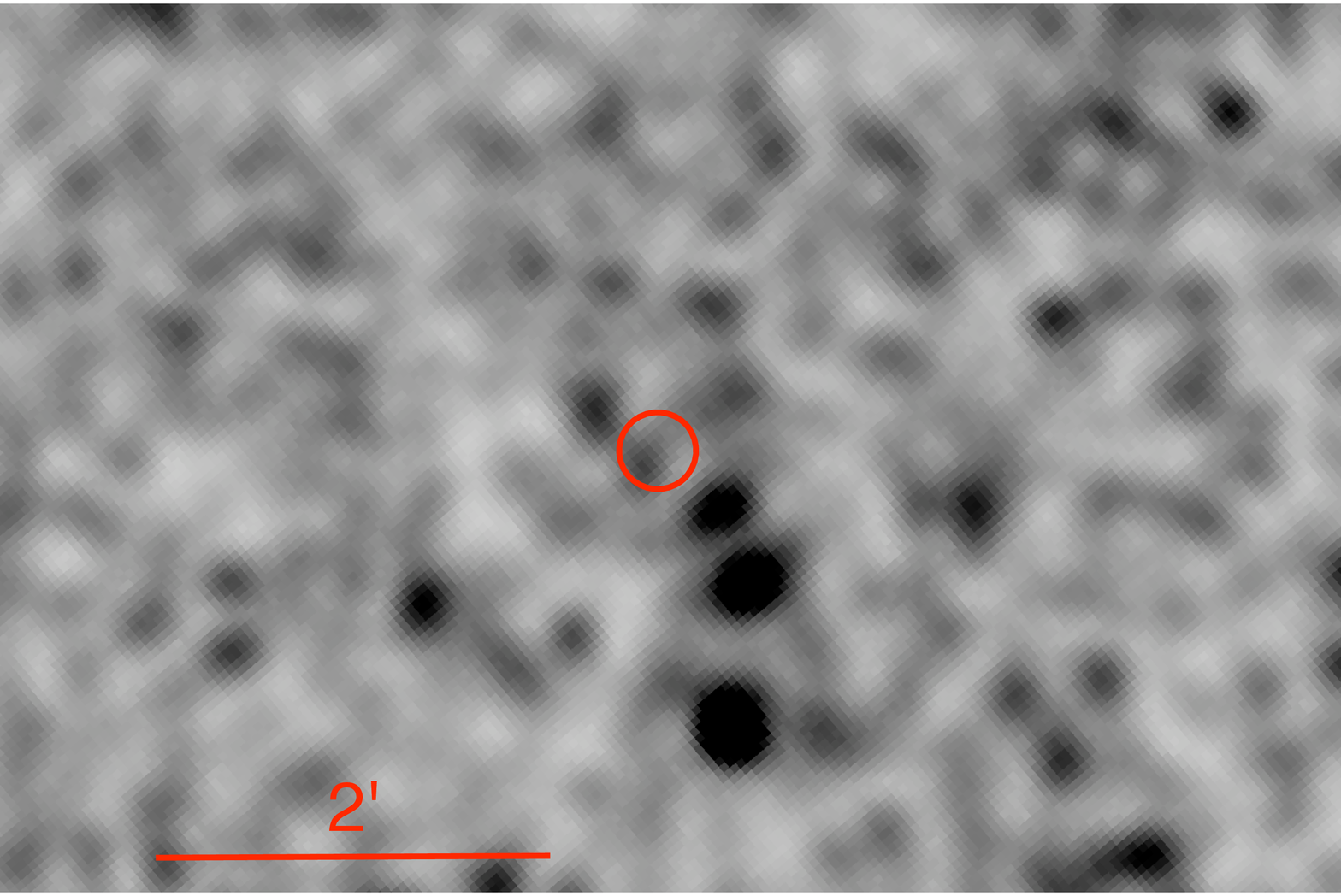}
\caption{\label{stack} A stacked image for 12 SLSN-I observed with the {\em Swift} XRT. The red circle marks the
position of the SNe in the stacked image. No source is clearly detected at this location, although numerous other
sources can be seen. The limit on any emission associated with the summed image is $L_X < 9 \times 10^{41}$ ergs s$^{-1}$ suggesting
that the mean luminosity for these SLSN is not in excess of $7 \times 10^{40}$ ergs s$^{-1}$. }
\end{center}
\end{figure}

A final possible origin for the X-ray emission is that it is powered by a relativistic jet from the compact object, but one that punctures the stellar envelope, similar to those that operate in gamma-ray bursts. This has some appeal, since GRBs can naturally create X-ray emission of high luminosity and since absorption by the ejecta is irrelevant.  However, as shown in Figure~\ref{lccomp} the X-ray emission from SCP 06F6 is an order of magnitude brighter than even luminous GRBs at the same epoch. This implies that if SCP 06F6 is an engine driven explosion, then it either lies at the extreme end of the population, or that the GRB-like event was not simultaneous.  Whether it is possible for a jet to break out of the stellar envelope on such a long timescale, despite the many instabilities and sources magnetic dissipation within the nebula (e.g., \citealt{Porth+12}), is also theoretically unclear. 

A requirement of both the `Exposed Nebula' or `Jet Break-Out' scenarios is that the central engine must be active at very late times after the core collapse.  This is possible in the case of a millisecond magnetar, since the duration of peak luminosity is set by the electromagnetic spin-down time, which can in principle be quite long depending on the initial rotation rate and dipole magnetic field strength of the magnetar (e.g.~\citealt{Metzger+11}).  A popular model for classical long duration GRBs posits that they are powered by rapid accretion onto a newly formed black hole. In most such models studied to date, the black hole accretes a compact disc created by material immediately exterior to innermost stable orbit.   Because of the small disk radius, the lifetime of the central engine is relatively short, and while some bursts show evidence for ongoing engine activity, this is generally over within a few thousands seconds of the burst trigger. More recently, some authors have focused on the possibility of discs created from material with higher angular momentum, as would originate from either extremely rapidly spinning, or from giant, stars \citep{quataert12,woosley12}.  In this case the engine could in principle be maintained for much longer if matter is indeed able to accrete efficiently from such large radii (although see \citealt{Fernandez&Metzger12}), in which case such events might power very long and highly luminous outbursts.  

These models were substantially developed in an attempt to explain the properties of {\em Swift} J1644+57 \citep{levan11}, an event that is now considered most likely to arise due to a relativistic variant of a tidal disruption event \citep{levan11,bloom11,burrows11,zauderer11}, in which a star
is disrupted by the strong tidal field of a supermassive black hole. However, it is interesting to note that the luminosity, and variability of {\em Swift} 1644+57 are very similar to those implied for SCP 06F6, and are also seen in other events such as {\em Swift} J2058+0516 \citep{cenko12}, and SDSS J120136.02+300305.5 \citep{saxton12}. The non-detection of SCP 06F6 by {\em Chandra} is marginally consistent with the intrinsic variability in {\em Swift} 1644+57, although an interesting possibility in a collapsar like model is that the material in the disc becomes depleted at late times.  At this point the engine itself may switch off, and could provide a natural explanation for the lightcurve of SCP 06F6. 

If SCP 06F6 were an engine driven explosion it is reasonable to consider whether the X-ray emission implies that the engine output were beamed in our direction. If this is the case, it is possible that
a GRB could have been observed from SCP 06F6 at the time of its core collapse.
It is interesting to note that one suggested GRB/SN association (albeit one based on
large error radii from BATSE) is than of SN~1997cy with GRB 970514 \citep{1997cy}. SN~1997cy was
one of the first identified SLSN (although it has also been suggested more
recently that this was a mis-identified SN Ia, \citealt{galyam12}).

We have examined {\em Swift} and Interplanetary Network (IPN) reported GRBs
for the period November to June 2006, however none of these is spatially coincident with SCP 06F6, implying
that it was not associated with a bright GRB beamed in our direction. An alternative
is that the GRB was not originally jetted at us, but that at late times we can observe
X-ray's from the source due to the lateral spreading of the jet. However, the X-ray luminosity
appears high for GRBs at such late times after the explosion (Figure~\ref{lccomp}), again suggesting that we are not witnessing a typical GRB-like event,  
unless the GRB occurred significantly after the core collapse \citep[e.g.][]{vietri00},
a model disfavoured by observations of GRB-SNe \citep[e.g.][]{hjorth12}. 

\section{Conclusions}

We have presented a analysis of X-ray emission from superluminous supernovae. 
Concluding that in one case, that of SCP 06F6, the X-ray emission at late times had a luminosity of $L_X \sim 10^{45}$ ergs s$^{-1}$. 
Such emission is too bright to easily be explained by a circumstellar interaction, unless the parameters are
rather fine tuned. 
The alternative model in which we observe a central engine in operation has appeal as a means
of readily achieving the X-ray luminosity, however it remains unclear if such models can
power the activity for long enough, and provide an explanation of the variability of the source. 
If the X-ray ray emission is produced by an engine within the SN it would
imply that SLSNe can create compact objects, and that pair instability events
in which the stellar core is completely destroyed likely cannot explain 
the observations. 

To make progress will require further observations, to determine which SLSNe produce luminous X-ray emission, and
to characterise the evolution of that emission in detail. 


\section*{Acknowledgements}
We thank the referee for a constructive and timely report which enhanced the clarity and quality of the paper. 
AJL, PJW, AMR acknowledge support from STFC, UK. Based on observations obtained with XMM-Newton, an ESA science mission
with instruments and contributions directly funded by
ESA Member States and NASA. This work made use of data supplied by the UK Swift Science Data Centre at the University of Leicester.

%
%

%
%

\begin{deluxetable}{lllll} 
\tablecolumns{8} 
\tablewidth{0pc}
\tablecaption{Log of X-ray observations of type I SLSN} 
\tablehead{ 
\colhead{Date-Obs} & \colhead{exposure (s)} & \colhead{$\Delta T$} & \colhead{F$_X$ (ergs cm$^{-2}$ s$^{-1}$)} &\colhead{L$_X$ (ergs s$^{-1}$)}}
\startdata
\hline
{\bf SCP 06F6} & $z=1.189$ & & \cite{barbary09} & \cite{quimby11}  \\
\hline
2006-08-02 & 9916 (MOS1) & 162  &$\approx 1 \times 10^{-13}$ & $\approx 1 \times 10^{45}$ \\
2006-08-02 & 8220 (MOS2)&  162 &$\approx 1 \times 10^{-13}$ &  $\approx 1 \times 10^{45}$\\
2006-08-02 & 2622  (pn) & 162  & $\approx 1 \times 10^{-13}$ &   $\approx 1 \times 10^{45}$\\
2006-11-04 & 4780 (ACIS) &256& $<1.4 \times 10^{-14}$ & $< 2.5 \times 10^{44}$ \\
\hline
{\bf PTF09atu} & $z=0.501$  & &  \cite{quimby11}\\
\hline
2009-08-18 & 4930 & 45 & $<8.1 \times 10^{-14}$ & $<7.3 \times 10^{43}$ \\
\hline
{\bf PTF09cnd} & $z=0.258$ & &  \cite{quimby11}\\
\hline
2009-08-18 & 3493 &36  & $<5.6 \times 10^{-14}$ & $<1.1 \times 10^{43}$ \\
2009-08-22 & 3555 & 40  & $<5.1 \times 10^{-14}$ & $<1.0 \times 10^{43}$\\
2009-08-26 & 3440 & 44 & $<6.6 \times 10^{-14}$  & $<1.3 \times 10^{43}$ \\
2009-08-30 & 4084 & 48  & $<5.7 \times 10^{-14}$ & $<1.1 \times 10^{43}$ \\
2009-09-03 & 2482 & 52 & $<7.3 \times 10^{-14}$ & $<1.4\times 10^{43}$  \\
2009-09-10 & 3059 & 59 & $<6.1 \times 10^{-14}$ & $<1.2 \times 10^{43}$\\
2009-09-23 & 2038 & 72& $<1.5 \times 10^{-13}$ &$<2.2 \times 10^{43}$ \\
2009-10-03 & 1906 & 82 & $<1.0 \times 10^{-13}$& $<2.0 \times 10^{43}$ \\
\hline
combined & 23961 & & $<1.2 \times 10^{-14}$ & $<2.5 \times 10^{42}$ \\
\hline
{\bf SN2009jh(PTF09cwl)} & $z=0.349$ &  & \cite{quimby11}\\
\hline
2009-09-04 &  3587 & 68 & $<1.1 \times 10^{-13}$ & $<4.5 \times 10^{43}$  \\
\hline
\bf{2010gx(PTF10cwr)} & $z=0.230$ & &  \cite{quimby11}\\
\hline
2010-03-19 & 3545 & 14& $<9.2 \times 10^{-14}$ & $<1.7 \times 10^{43}$ \\
2010-03-20 & 990 &  15 & $<2.7 \times 10^{-13}$ & $<5.6  \times 10^{43}$\\
2010-03-25 & 2121 &20 &$<1.5 \times 10^{-13}$ & $<2.2 \times 10^{43}$\\
2010-03-31 & 1790 &26 & $<2.3 \times 10^{-13}$ & $<3.5 \times 10^{43}$\\
2010-04-08 & 2206 & 34& $<1.9 \times 10^{-13}$ & $<2.8 \times 10^{43}$\\
2010-04-16 & 1459 &42 & $<2.0 \times 10^{-13}$ & $<3.1 \times 10^{43}$ \\
2010-04-23 & 1356 &49 & $<2.1 \times 10^{-13}$ & $<3.3 \times 10^{43}$\\
2010-05-02 & 2224 &58 & $<1.3 \times 10^{-13}$ &$<2.1 \times 10^{43}$\\
2010-05-08 & 1928 &64 & $<2.4 \times 10^{-13}$ & $<3.8  \times 10^{43}$\\
2010-05-13 & 2086  & 69& $<1.4 \times 10^{-13}$ & $<2.1 \times 10^{43}$ \\
\hline
combined & 19707 &  & $<3.0 \times 10^{-14}$ & $<4.6 \times 10^{42}$ \\
\hline
{\bf PTF10hgi} & $z \sim 0.1$ &  & \cite{atel2740}\\
\hline
2010-07-13 & 1718 & 59  & $< 1.9 \times 10^{-13}$ & $4.5 \times 10^{42}$ \\
2010-07-18 & 2836 & 64 & $< 1.1 \times 10^{-13}$ & $2.8 \times 10^{42}$\\
\hline
combined & 4554 & - &  $< 7.0 \times 10^{-14}$ & $1.7 \times 10^{42}$\\
\hline
{\bf SN 2010kd} & $z=0.101$  & & \cite{vinko10,vinko12} \\
\hline
2010-11-30 & 1785 & 16  & $< 2.2 \times 10^{-13}$ & $5.4 \times 10^{42}$\\
2010-11-30 &  1780 & 16 & $< 2.3 \times 10^{-13}$ & $5.6 \times 10^{42}$\\
2010-12-16 &  3177 & 32 & $< 8.5 \times 10^{-14}$ & $2.1 \times 10^{42}$\\
2010-12-19 &  2808 & 35 & $< 1.2 \times 10^{-13}$ & $3.1 \times 10^{42}$\\
2010-12-22 &  3470 & 38 & $< 8.9 \times 10^{-14}$ & $2.2 \times 10^{42}$\\
2010-12-25 &  3443 & 41 & $< 7.9 \times 10^{-14}$ & $1.9 \times 10^{42}$ \\
\hline
combined & 16463 & -   & $< 2.7 \times 10^{-14}$ & $6.5 \times 10^{41}$ \\
\hline
{\bf PTF11dij (CSS110406)} &  $z=0.143$  & & \cite{atel3343} & \cite{atel3344}  \\
\hline
2011-05-14 & 2971 &45 &$<1.3 \times 10^{-13}$ & $<7.0 \times 10^{42}$  \\
2011-05-30 & 4027  &61 &$<6.5 \times 10^{-14}$ & $<3.5 \times 10^{42}$ \\
2011-06-06  & 1018 &68 &$<2.5 \times 10^{-13}$ & $<1.3 \times 10^{43}$ \\
2011-06-07  & 1783 &69 &$<1.4 \times 10^{-13}$ & $<7.6 \times 10^{42}$ \\
2011-06-08 & 1457 & 70& $<1.8 \times 10^{-13}$ & $<1.0 \times 10^{43}$\\
2012-03-11 & 953 &347 &$<3.3 \times 10^{-13}$ & $<1.7 \times 10^{43}$ \\
2012-03-14 & 401 &350 & $<7.3 \times 10^{-13}$ & $<4.5 \times 10^{43}$\\
2012-03-18 & 1033 &354 &$<3.9 \times 10^{-13}$ & $<5.3 \times 10^{43}$\\
2012-03-22 & 431 &358 &$<9.5 \times 10^{-13}$ & $<5.0 \times 10^{43}$ \\
2012-03-31 & 484 &367 &$<3.5\times 10^{-13}$ & $<3.1 \times 10^{43}$\\
2012-04-03 & 474 &370 & $<5.4 \times 10^{-13}$ & $<2.9 \times 10^{43}$\\
2012-04-12 & 140 &379 &$<1.8 \times 10^{-12}$ & $<1.0 \times 10^{44}$ \\
\hline
combined & 15200 & & $<2.4 \times 10^{-14}$ & $<1.2 \times 10^{42}$ \\
\hline
{\bf PTF11dsf} & $z=0.385$ & & \cite{atel3465} \\
\hline
2011-06-03 & 3766.3 &  22 & $<8.6 \times 10^{-14}$ & $<4.4 \times 10^{43}$ \\
\hline
{\bf PTF11rks} & $z=0.19$ &  &   \cite{atel3841} \\
\hline
2011-12-30 & 3473 &9 & $<2.2 \times 10^{-13}$  & $<2.2 \times 10^{43}$\\
2012-01-01 & 4295 &11 &$<7.7 \times 10^{-14}$   & $<7.7 \times 10^{42}$\\
2012-01-05 & 4182 &15 &$<1.0 \times 10^{-13}$  & $<1.0 \times 10^{43}$\\
2012-01-10 & 4939 & 20&$<6.2 \times 10^{-14}$  & $<6.2 \times 10^{42}$\\
2012-01-15 & 4546 & 25 & $<7.9 \times 10^{-14}$ &$<7.9 \times 10^{42}$ \\
\hline 
combined & 21434.9 & &  $<2.2 \times 10^{-14}$ & $<2.2 \times 10^{42}$\\
\hline
{\bf PS1-12fo(CSS120121)} & $z=0.175$ & &\cite{atel3873} & \cite{atel3918} \\
\hline
2012-02-13 & 4759 & 44 & $< 1.0 \times 10^{-13}$ & $<8.7 \times 10^{42}$   \\
\hline
{\bf PTF12dam} &$z=0.107$ & &  \cite{atel4121}  \\
\hline
2012-05-22 & 2460 & 32 & $< 1.1 \times 10^{-13}$ & $2.9 \times 10^{42}$\\
2012-05-30 & 1991 & 40 & $< 1.3 \times 10^{-13}$ & $3.5 \times 10^{42}$\\
2012-06-07 & 1314 & 48 & $< 1.9 \times 10^{-13}$ & $5.3 \times 10^{42}$\\
2012-06-13 & 1615 & 54 & $< 1.6 \times 10^{-13}$ & $4.3 \times 10^{42}$\\ 
2012-06-20 & 1800 & 61 & $< 1.4 \times 10^{-13}$ & $3.9 \times 10^{42}$ \\
2012-06-27 & 930 & 68  & $< 2.7 \times 10^{-13}$ & $7.5 \times 10^{42}$\\
2012-06-28 & 915 & 69 & $< 2.7 \times 10^{-13}$ & $7.6 \times 10^{42}$\\
2012-07-04 & 1058 & 75 & $< 3.4 \times 10^{-13}$ & $9.3 \times 10^{42}$\\
2012-07-11 & 2560 & 82 & $< 9.8 \times 10^{-14}$ & $2.7 \times 10^{42}$\\
2012-07-18 & 1173 & 89 & $< 2.6 \times 10^{-13}$ & $6.6 \times 10^{42}$\\
\hline
combined & 15816 & - &  $< 2.3 \times 10^{-14}$ & $6.3 \times 10^{41}$\\
\hline
{\bf CSS121015} & $z=0.286$ & & \cite{atel4498} & \cite{atel4512} \\
\hline
2012-10-24 & 3099 & 9 & $< 1.3 \times 10^{-13}$ & $<3.2 \times 10^{43}$ \\
2012-10-25 & 3839 & 10 &  $< 1.2 \times 10^{-13}$ & $<2.8 \times 10^{43}$\\
2012-10-26 & 2342 & 11& $< 1.4 \times 10^{-13}$ & $<3.5\times 10^{43}$\\
2012-10-31 & 2803 & 16 & $< 1.2 \times 10^{-13}$ & $<2.9 \times 10^{43}$ \\
2012-11-02 & 559 & 18 & $< 7.9 \times 10^{-13}$ & $<2.0 \times 10^{44}$ \\
2012-11-04 & 3671 & 20 & $< 8.5 \times 10^{-14}$ & $<2.1 \times 10^{43}$ \\
2012-11-06 &  3884 & 22 & $< 8.1 \times 10^{-14}$ & $<1.8 \times 10^{43}$ \\
2012-11-08 & 4245 & 24 & $< 7.3 \times 10^{-14}$ & $<8.8 \times 10^{42}$ \\
2012-11-09 & 216 & 25 & $< 1.4 \times 10^{-12}$ & $<3.6 \times 10^{44}$ \\
2012-11-12 &4122 & 28 & $< 1.4 \times 10^{-12}$ & $<3.5 \times 10^{43}$ \\
2012-11-14 & 4278 & 30 & $< 1.3 \times 10^{-13}$ & $<3.4 \times 10^{43}$ \\
2012-11-16 & 2079 & 32 & $< 1.6 \times 10^{-13}$ & $<4.0 \times 10^{43}$ \\
\hline
combined & 35135 & - &   $1.3^{+0.9}_{-0.9} \times 10^{-14}$ & $3.0^{+2.4}_{-2.3} \times 10^{42}$\\
\hline
\enddata
\tablecomments{X-ray observations of known SNSN Ic. For each observation we give fluxes and luminosities, corrected
for galactic foreground $N_H$ over the 0.2-10 keV band for a source with photon index $\Gamma=2$ (with the exception
of SCP 06F6, see section~\ref{scp06f6}. The observation
dates are shown as well as the time since the discovery of the SN (note that this is not necessarily close to the time of
core collapse). These dates are 21 February 2006 (SCP 06F6), 4 July 2009 (PTF09atu), 13 July 2009 (PTF09cnd), 28 June 2009 (SN 2009jh/PTF09cwl), 
5 March 2010 (SN 2010gx/PTF10cwr), 15 May 2010 (PTF10hgi), 14 November 2010 (2010kd), 30 March 2011 (PTF11dij), 12 May 2011 (PTF11dsf), 21 Dec 2011 (PTF11rks),
31 Dec 2011 (PS1-12fo), 20 April 2012 (PTF12dam), 15 Oct
2012 (CSS121015). Note that formally CSS121015 is detected in the stacked image at $\sim 99$\% confidence. This may be a genuine detection, but given the number of measurements made
this is consistent with random fluctuations. }
\label{x-ray}
\end{deluxetable}

\end{document}